\documentclass[aps,prl,twocolumn,amsmath,amssymb,nofootinbib,superscriptaddress]{revtex4-1}
\usepackage{times}
\usepackage[pdftex]{graphicx}
\usepackage{dcolumn}
\usepackage{bm}
\usepackage{amsmath}
\usepackage{indentfirst}
\usepackage{float}
\usepackage[colorlinks]{hyperref}
\usepackage[dvipsnames]{xcolor}

\usepackage{makecell}
\usepackage{diagbox}

\newcommand{\nodes}{\mathcal{V}}
\newcommand{\edges}{\mathcal{E}}

\newcommand{\graph}{\mathcal{G}}
\newcommand{\cpath}{\mathcal{P}}

\newcommand{\edgemax}{\mathcal{M}}

\newcommand{\fidelity}{\mathcal{F}}

\begin{document}

\title{Verifying Random Quantum Circuits with Arbitrary Geometry Using Tensor Network States Algorithm}

\author{Chu Guo}
\thanks{These authors contribute equally to this work}
\email{guochu604b@gmail.com}
\affiliation{Key Laboratory of Low-Dimensional Quantum Structures and Quantum Control of Ministry of Education, Department of Physics and Synergetic Innovation Center for Quantum Effects and Applications, Hunan Normal University, Changsha 410081, China}

\author{Youwei Zhao}
\thanks{These authors contribute equally to this work.}
\affiliation{Hefei National Laboratory for Physical Sciences at the Microscale and Department of Modern Physics, University of Science and Technology of China, Hefei 230026, China}
\affiliation{Shanghai Branch, CAS Center for Excellence in Quantum Information and Quantum Physics, University of Science and Technology of China, Shanghai 201315, China}
\affiliation{Shanghai Research Center for Quantum Sciences, Shanghai 201315, China}

\author{He-Liang Huang}
\email{quanhhl@ustc.edu.cn}
\affiliation{Henan Key Laboratory of Quantum Information and Cryptography, Zhengzhou, Henan 450000, China}
\affiliation{Hefei National Laboratory for Physical Sciences at the Microscale and Department of Modern Physics, University of Science and Technology of China, Hefei 230026, China}
\affiliation{Shanghai Branch, CAS Center for Excellence in Quantum Information and Quantum Physics, University of Science and Technology of China, Shanghai 201315, China}
\affiliation{Shanghai Research Center for Quantum Sciences, Shanghai 201315, China}

\begin{abstract}
The ability to efficiently simulate random quantum circuits using a classical computer is increasingly important for developing Noisy Intermediate-Scale Quantum devices. Here we present a tensor network states based algorithm specifically designed to compute amplitudes for random quantum circuits with arbitrary geometry. Singular value decomposition based compression together with a two-sided circuit evolution algorithm are used to further compress the resulting tensor network. To further accelerate the simulation, we also propose a heuristic algorithm to compute the optimal tensor contraction path. We demonstrate that our algorithm is up to $2$ orders of magnitudes faster than the Sch$\ddot{\text{o}}$dinger-Feynman algorithm for verifying random quantum circuits on the $53$-qubit Sycamore processor, with circuit depths below $12$. We also simulate larger random quantum circuits up to $104$ qubits, showing that this algorithm is an ideal tool to verify relatively shallow quantum circuits on near-term quantum computers.
\end{abstract}

\date{\today}
\maketitle

\address{}

\vspace{8mm}


Recent progress of quantum computing hardware has achieved more than $50$ qubits with gate operation fidelities higher than $99 \%$, marking the entering of the Noisy Intermediate-Scale Quantum (NISQ) computing era~\cite{AruteMartinisQuantumSupremacy2019, preskill2018quantum, huang2020superconducting}. Accompanying with the hardware progresses, there is a stimulated interest in exploring suitable near-term applications for such devices~\cite{Google2020a,Google2020b,HuangPan2020,liu2019hybrid,havlivcek2019supervised,kandala2017hardware,kokail2019self,cong2019quantum,hempel2018quantum}. A central difficulty when building NISQ hardwares with even more qubits is to maintain the high qualities of the devices as to the quantum gate operations as well as the quantum measurements. Thus efficient ways to benchmark quantum hardwares become increasingly important since it enables researchers and engineers to rapidly evaluate the performance of the quantum processors and continuously improve them.




Randomized benchmarking has been a standard tool to benchmark quantum gate operations~\cite{KnillWineland2008, emerson2005scalable}. However, it is difficult to be scaled up to quantum circuits with several tens of qubits due to the rapid growth of complexity. In Ref.~\cite{BoixoNeven2017}, random quantum circuits (RQCs) were proposed to benchmark the performance of quantum computing devices. RQC possesses at least two important features which make it an ideal problem for NISQ hardwares to solve: 1) RQCs often consist of interlacing layers of single- and nearest-neighbour two-qubit gate operations which are extremely friendly for current quantum computing hardwares and 2) with a total of only several hundreds of two-qubit gate operations, RQCs could already generate highly entangled quantum states which are extremely hard to reproduce with even the best supercomputers~\cite{bouland2019complexity,aaronson2016complexity,bremner2016average}. For those reasons, RQCs have been employed to demonstrate \textit{quantum supremacy}~\cite{AruteMartinisQuantumSupremacy2019, harrow2017quantum,neill2018blueprint}. An important ingredient when using RQCs to benchmark NISQ hardwares is to simulate RQCs with the best classical algorithm, which can server as 1) a baseline for the classical complexity of the problem and 2) a verification tool for the outputs of the quantum devices. However, verifying the 53-qubit RQC reported in Ref.~\cite{AruteMartinisQuantumSupremacy2019} has already used $1$ million cores for $5$ hours, which poses a huge challenge for the verification of RQCs on large-scale quantum computing hardwares in the next stage.

So far, various classical algorithms have been proposed to simulate RQCs. Depending on the way that the quantum state is represented, those algorithms can roughly be divided into three categories: 1) directly storing and evolving the quantum state~\cite{DeIto2007,SmelyanskiyGuzik2016,HanerSteiger2017,PednaultWisnieff2017}; 2) tensor network contraction based methods, where the quantum state and the quantum circuit are treated altogether as a large tensor network, and then amplitudes are obtained by contracting this tensor network with certain contraction path~\cite{MarkovShi2008,BoixoNeven2017b,ChenGuo2018,LiYang2018,ChenShi2018,VillalongaMandra2018,VillalongaMandra2019} and 3) tensor network states (TNS) based methods~\cite{MccaskeyHumble2018,GuoWu2019}, adapted from the tensor network states algorithm originally developed in quantum many-body physics~\cite{Schollwock2011,VerstraeteCirac2004,VerstraeteCirac2006}. The first category is limited by the memory, since current most powerful supercomputers can only store about 50 qubits. The major differences between the second and last categories are that in tensor network states based methods: i) the quantum states are directly stored as tensor networks and the gate operations are applied onto those tensor networks subsequently, as a result quantum measurements can be simulated straightforwardly in addition to computing the amplitudes, ii) there is in general a compression stage following each two-qubit gate operation.

TNS based methods are often designed for regular lattices, such as a one-dimensional lattice or a square lattice, which may not be easily adapted to the topology of current NISQ hardwares. Moreover, there may exist some bad qubits inside the lattice which are not used for the computation at all~\cite{AruteMartinisQuantumSupremacy2019}. In this work, we present a TNS based algorithm which would be suitable for arbitrary lattice geometry. For the specific task of computing a single amplitude, we use a two-sided circuit evolution technique which could maximumly compress the size of the resulting tensor network, accompanied with a heuristic algorithm used to search for the optimal tensor contraction path.
We demonstrate the efficiency of this method by applying it to simulate $53$-qubit RQCs up to a depth of 11, and comparing its performance with the Sch$\ddot{\text{o}}$dinger-Feynman algorithm~\cite{MarkovBoixo2018}, which was used as the benchmarking baseline for demonstrating quantum supremacy~\cite{AruteMartinisQuantumSupremacy2019}.
We also apply our algorithm to study large quantum circuits, showing that it is an ideal tool for fast verification of relatively shallow RQCs running on NISQ hardwares.



\begin{figure}[tbp]
\includegraphics[width=\columnwidth]{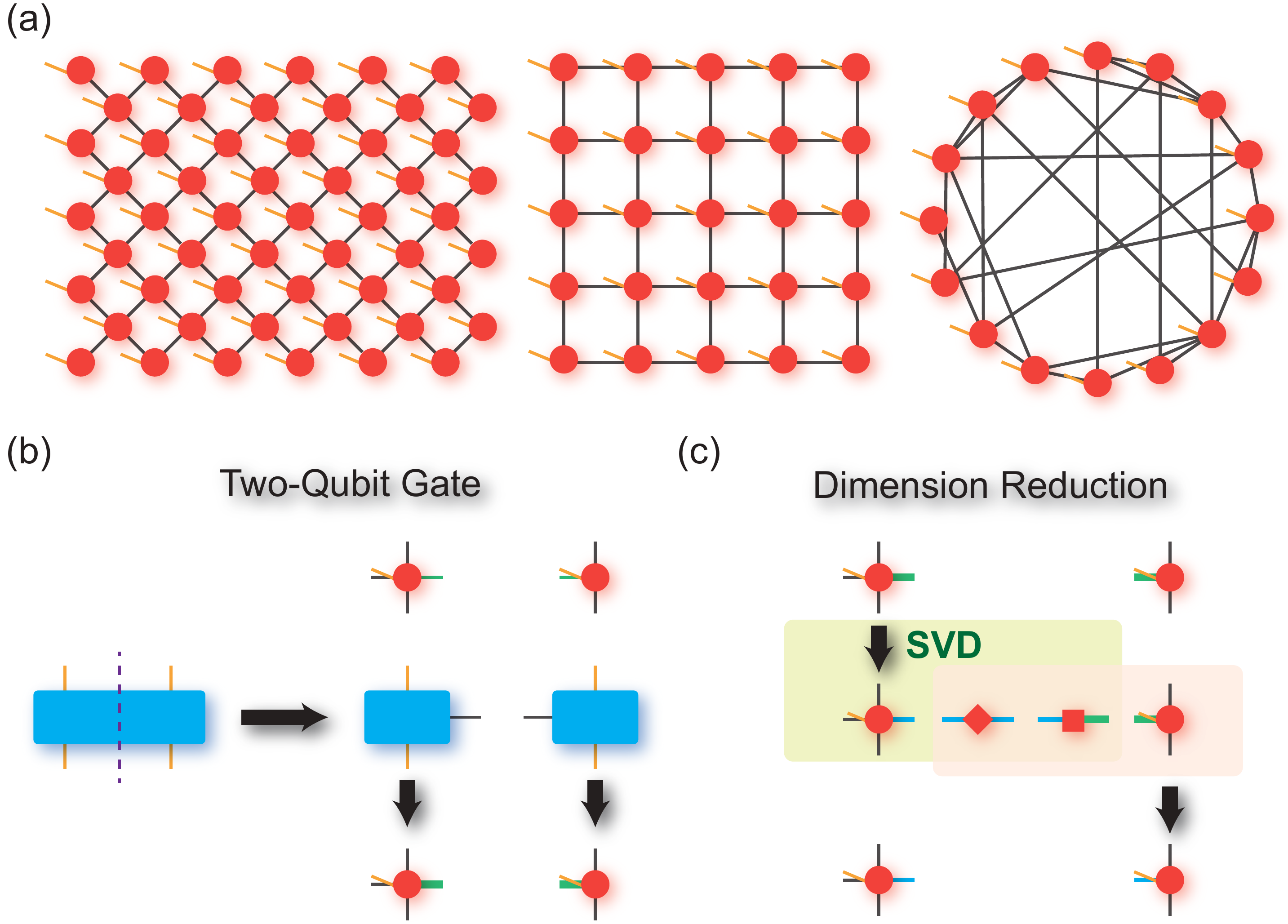}
\caption{(a) Three examples of graph geometries: the Sycamore processor, a square lattice as well as an arbitrarily connected lattice from left to right. Each red circle stands for one qubit, and the black lines means the connections between the qubits, which also correspond to the auxiliary indexes of the tensor network states. The orange lines represent the physical indexes. (b) Two-qubit gate operation. The left rectangle with four orange lines and a dashed cut in between represents the splitting of a two-qubit gate into two three dimensional tensors as in Eq.(\ref{eq:gatesplit}). The block on the right hand side of the black arrow shows the procedure of the two-qubit gate operation, which corresponds to Eqs.(\ref{eq:result1}, \ref{eq:result2}). We have used a thicker green line to explicitly indicate that the size of the auxiliary index increases after the two-qubit gate operation. (c) SVD compression of the resulting tensors from two-qubit gate operation corresponding to Eqs.(\ref{eq:svd}, \ref{eq:svd2}).}   \label{fig:fig1}
\end{figure}

\textit{State initialization and gate operations.} We assume that the lattice geometry can be represented by a connected graph, where each node represents a qubit and each edge means that there is at least one two-qubit gate applied on the two qubits connected by this edge. We denotes the graph as $\graph=\{\nodes, \edges\}$, where $\nodes$ represents the nodes (qubits) and $\edges$ represents the edges. We use $\edges_j$ to denote all the edges connected to the $j$-th node $\nodes_j$. The quantum state on such a graph $\graph$ is initialized as a tensor network state as follows. For each node $\nodes_j$ with $\dim(\edges_j)$ edges, we initialize a $\dim(\edges_j)+1$ dimensional tensor $A^{\sigma_j}_{a_1^j a_2^j \dots a^j_{\dim(\edges_j)}}$ (we will simply denote it as $A^j$ for short if the details of the indexes are not important in the context) of size $2\times 1\times\dots \times1$, reshaped from the vector $[1, 0]$ ($[0, 1]$) corresponding to the single-qubit state $\vert 0\rangle$ ($\vert 1\rangle$). The first index is the physical index and the rest indexes are the auxiliary indexes. Moreover, if two nodes $\nodes_k$ and $\nodes_l$ are connected by an edge, then one of the auxiliary index of $A^k$, say $a^k_m$, should be contracted with one of the auxiliary index of $A^l$, say $a^l_n$ and we would simply say that those two auxiliary indexes $a^k_m$ and $a^l_n$ are connected. The initial $N$-qubit quantum state $\vert 0\rangle^N$ is thus written as a tensor network
\begin{align}\label{eq:initialstate}
\vert 0\rangle^N = \mathcal{F}(A^{\sigma_1=0}_{a_1^1 \dots a^1_{\dim(\edges_1)}}\dots A^{\sigma_N=0}_{a_1^N\dots a^N_{\dim(\edges_N)}}),
\end{align}
where we have written $\sigma_j=0$ on the superscript of each tensor to explicitly indicate that $A^{\sigma_j}_{a_1^j\dots a^j_{\dim(\edges_j)}} = 1$ for $\sigma_j=0$ and $0$ otherwise. $\mathcal{F}$ means to contract all the pairs of connected auxiliary indexes.
The central difference of Eq.(\ref{eq:initialstate}) from a tensor network state on a regular lattice is that the number of auxiliary indexes of each node is not a constant, but is determined by the graph topology, or more precisely the number of edges connected to it. In Fig.~\ref{fig:fig1}(a) we show several possible graph geometries, and the corresponding TNS.

For a two-qubit gate denoted as $O_{\sigma_k\sigma_l}^{\sigma_k'\sigma_l'}$ acting on the $k$- and $l$-th qubits, we first split it into two three-dimensional tensors using singular value decomposition (SVD) as in~\cite{GuoWu2019}
\begin{align}\label{eq:gatesplit}
O_{\sigma_k\sigma_l}^{\sigma_k'\sigma_l'} = \sum_{s=1}^{\chi_o} P_{\sigma_k s}^{\sigma_k'} Q_{\sigma_l s}^{\sigma_l'},
\end{align}
where the size of the auxiliary index $s$ is equal to the number of non-zero singular values, denoted as $\chi_o=\dim(s)$. For a controlled gate $\chi_o=2$ while for the iSWAP gate as well as the fSim gate used in Ref.~\cite{AruteMartinisQuantumSupremacy2019} $\chi_o = 4$. Assuming $A^k$ and $A^l$ are connected by two auxiliary indexes $a^k_m$ and $a^l_n$, then the two-qubit gate operation on $A^k$ and $A^l$ can be denoted as
\begin{align}
&A^{\sigma_k'}_{[a_1^k\dots a^k_{\dim(\edges_k)}]_m \mathbf{a}_m^k} \leftarrow \sum_{\sigma_k} P_{\sigma_k s}^{\sigma_k'} A^{\sigma_k}_{a_1^k\dots a^k_{\dim(\edges_k)}}; \label{eq:result1} \\
&A^{\sigma_l'}_{[a_1^l\dots a^l_{\dim(\edges_l)}]_n \mathbf{a}_n^l} \leftarrow \sum_{\sigma_l} Q_{\sigma_l s}^{\sigma_l'} A^{\sigma_l}_{a_1\dots a_{\dim(\edges_l)}}, \label{eq:result2}
\end{align}
where $[a_1^k\dots a^k_{\dim(\edges_k)}]_m = a_1^k\dots a_{m-1}^k a_{m+1}^k \dots a^k_{\dim(\edges_k)}$ and similarly for $[a_1^l\dots a^l_{\dim(\edges_l)}]_n$. We have also used $\mathbf{a}_m^k = a_m^k s$ and $\mathbf{a}_n^l = a_n^l s$, from which we can see clearly that after a two-qubit gate operation, the size of the auxiliary dimensions $a_m^k$ and $a_n^l$ are increased by a factor of $\chi_o$. The procedure of a two-qubit gate operation is also shown in Fig.~\ref{fig:fig1}(b). Single-qubit gates are not considered since they can be absorbed into two-qubit gates using gate fusion.


\textit{Compression by SVD.} As we have pointed out in the introduction, an important feature of TNS based algorithms is that the resulting tensors after each two-qubit gate operation will be compressed, which can be done as follows. First we perform SVD on one of the resulting tensors in Eqs.(\ref{eq:result1}, \ref{eq:result2}), say $A^{\sigma_k'}$, as
\begin{align}\label{eq:svd}
A^{\sigma_k'}_{[a_1^k\dots]_m \mathbf{a}_m^k} = \sum_{s' s''}U^{\sigma_k'}_{[a_1^k\dots]_m s'} S_{s' s''} V_{s'' \mathbf{a}_m^k},
\end{align}
where only the nonzero singular values of $S$ are kept. Then one absorbs the matrix $V_{s' \mathbf{a}_m^k}^{\prime} = \sum_{s''} S_{s' s''} V_{s'' \mathbf{a}_m^k}$ into the other tensor $A^{\sigma_l'}$ as
\begin{align}\label{eq:svd2}
A^{\sigma_l'}_{[a_1^l\dots]_n s'} \leftarrow \sum_{\mathbf{a}_m^k}A^{\sigma_l'}_{[a_1^l\dots]_n \mathbf{a}_m^k} V_{s' \mathbf{a}_m^k}^{\prime}.
\end{align}
Thus the size of the auxiliary index $a_m^k$ changes from $\dim(a_m^k)\chi_o$ to $\dim(s')$, satisfying $\dim(s')\leq \dim(a_m^k)\chi_o$, and similarly for $a_n^l$. The SVD compression procedure is shown in Fig.~\ref{fig:fig1}(c).

\begin{figure}[tbp]
\includegraphics[width=\columnwidth]{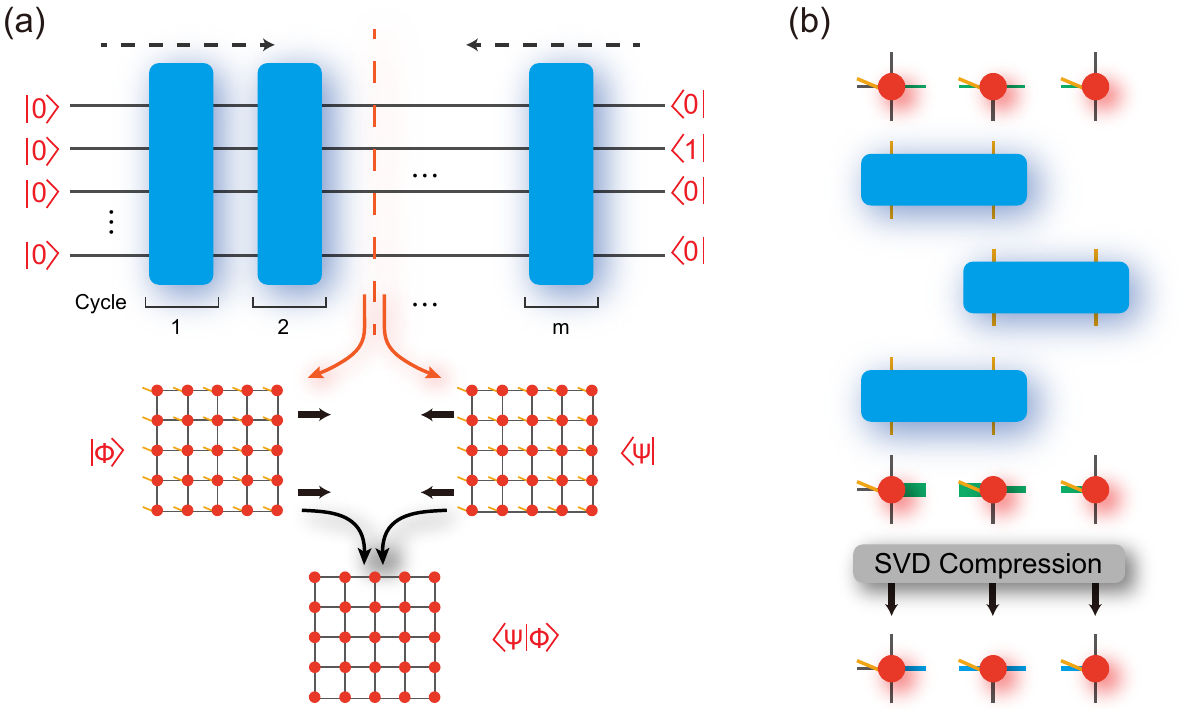}
\caption{(a) Two-sided circuit evolution. Each blue rectangle represents a group of two-qubit gates (one cycle). The $m$ cycles are further divided into two groups as indicated by the vertical red dashed line in the middle of the circuit. The left cycles are applied onto the initial state $\vert0\rangle^N$ from left to right, while the right cycles are applied inversely onto the target state $\vert 01\dots\rangle $ from right to left. The evolution results in two tensor network states as shown with the two lattices below the circuit. The physical indexes of those two tensor network states are then contracted, which results in the tensor network on the bottom with no open legs. (b) Automatic compression of the result tensors from the DCD pattern by SVD. Here each blue rectangle stands for a two-qubit gate.}   \label{fig:fig2}
\end{figure}

In the follow we identify two situations that we could have $\dim(s') < \dim(a_m^k)\chi_o$.
First, when the $k$-th qubit is applied on by a two-qubit gate $O_{\sigma_k\sigma_l}^{\sigma_k'\sigma_l'}$ with $\chi_o=4$ for the first time, we will have from Eq.(\ref{eq:svd}) that
\begin{align}\label{eq:shouwei}
\dim(s')\leq \min(2\dim([a_1^k\dots]_m), \dim(\mathbf{a}_m^k))=2.
\end{align}
Namely the size of the corresponding auxiliary index can at most increase to $2$.
We note that it is pointed out in Ref.~\cite{AruteMartinisQuantumSupremacy2019} that the fSim gate in the first two layers can be simplified into a controlled phase gate with $\chi_o=2$, since it can be decomposed into a controlled phase gate and an iSWAP gate.
In contrast for our method the compression in Eq.(\ref{eq:shouwei}) naturally results from Eqs.(\ref{eq:svd}, \ref{eq:svd2}) for any two-qubit gate satisfying $\chi_o > 2$.
Till now such compressions are only possible in the first few layers of gate operations.
Now recalling that for the task of computing amplitudes, the quantum circuit starts from a separable quantum state corresponding to a bitstring $00\dots 0$ and is finally projected onto another separable quantum state corresponding to a bitstring $s_1s_2\dots s_N$ with $s_n=0,1$. To make use of the compression in Eq.(\ref{eq:shouwei}) also in the last layers of gate operations, we can divide the two-body gates into two groups and perform a \textit{two-sided circuit evolution}, that is, the first group of gates are applied onto the initial quantum state $\vert0\rangle^N$, while the second group of gates are applied inversely onto target quantum state $\vert s_1s_2\dots s_N\rangle$, then one obtains one amplitude by computing the overlap between two resulting TNS. This procedure is shown in Fig.~\ref{fig:fig2}(a).

In the second case, we consider the DCD pattern as described in Ref.~\cite{AruteMartinisQuantumSupremacy2019}, which means that there are three successive two-qubit gates acting on the qubit pairs $(k, l)$, $(l, r)$ and $(k,l)$. We look at the tensor $A^k$ and assume that its auxiliary index $a^k_m$ is connected to the tensor $A^l$. $A^k$ is applied on twice, therefore the size of $a^k_m$ would increase to $\dim(a_m^k)\chi_o^2$, while the sizes of the rest auxiliary indexes of $A^k$ remain unchanged. Moreover, DCD pattern often happens at the boundary, such that $A^k$ only has very few auxiliary indexes. As a result it is very likely that $\dim(a_m^k)\chi_o^2 > 2\dim([a_1^k\dots a^k_{\dim(\edges_k)}]_m)$, in which case the size of $a_m^k$ will get compressed by Eq.(\ref{eq:svd}) and thus grows slower than by a factor of $\chi_o^2$. The occurrence of this pattern as well as the compression of the resulting tensors are shown in Fig.~\ref{fig:fig2}(b). We note that this compression is done automatically by Eqs.(\ref{eq:svd}, \ref{eq:svd2}) without additional manual efforts.




\textit{Overlap between two tensor network states.} As shown in Fig.~\ref{fig:fig2}(a), the two-sided circuit evolution will result in two TNS corresponding to two output quantum states $\vert \phi\rangle$ and $\langle \psi\vert$ respectively. Writing $\vert \phi\rangle = \mathcal{F}(A^{\sigma_1}_{a_1^1 \dots a^1_{\dim(\edges_1)}}\dots A^{\sigma_N}_{a_1^N\dots a^N_{\dim(\edges_N)}})$ and $\langle \psi\vert = \mathcal{F}(B^{\sigma_1}_{b_1^1 \dots b^1_{\dim(\edges_1)}}\dots B^{\sigma_N}_{b_1^N\dots b^N_{\dim(\edges_N)}})$, the overlap of $\vert \phi\rangle$ and $\langle \psi\vert$ can be computed by contracting all the physical indexes between them, that is,
\begin{align} \label{eq:overlap}
\langle \psi \vert  \phi\rangle = \mathcal{F}(C_{c_1^1 \dots c^1_{\dim(\edges_1)}}\dots C_{c_1^N\dots c^N_{\dim(\edges_N)}}),
\end{align}
where $C_{c_1^l\dots c^l_{\dim(\edges_j)}} = \sum_{\sigma_l} A^{\sigma_l}_{a_1^l \dots a^l_{\dim(\edges_l)}} B^{\sigma_l}_{b_1^l \dots b^l_{\dim(\edges_l)}}$ for $1\leq l\leq N$. Eq.(\ref{eq:overlap}) is a tensor network on graph $\graph$. Directly contracting this tensor network will generally result in high-dimensional intermediate tensors which have to be stored distributedly, leading to cross-node data communication costs~\cite{GuoWu2019}. To overcome this difficulty, one can \textit{cut} a few legs in Eq.(\ref{eq:overlap}) as done in Ref.~\cite{VillalongaMandra2018}. For example, cutting the auxiliary dimension $c_m^k$ amounts to splitting the $m$-th leg of the tensor $C^k$ into $\dim(c_m^k)$ slices (the same for the tensor which connects to $C^k$ via $c_m^k$), as a result the tensor network in Eq.(\ref{eq:overlap}) is split into $\dim(c_m^k)$ sub tensor networks in which the auxiliary index $c_m^k$ is removed. Each sub tensor network produces a single scalar. Summing over these scalars results in the final amplitude.

In addition, we propose a heuristic algorithm to search for the optimal tensor contraction path. Based on the observation that current NISQ hardwares have a (quasi)-regular two-dimensional geometrical structure, we made three assumptions that an optimal tensor contraction path $\cpath_{i_1\dots i_N}$ needs to satisfy: 1) $\cpath_{i_1\dots i_N}$ starts from a qubit on the boundary; 2) the rank of the intermediate tensors appear along this path is bounded by a maximum value $\edgemax$; 3) For each $m$, the subgraph formed by the qubits $\{i_1,\dots, i_m\}$ is \textit{almost} connected. These assumptions allow us to neglect most of the paths. And we are able to come up with an efficient searching algorithm using state compression dynamical programming technique, which is detailed in the supplementary~\cite{supp}.


\begin{figure}[tbp]
\includegraphics[width=\columnwidth]{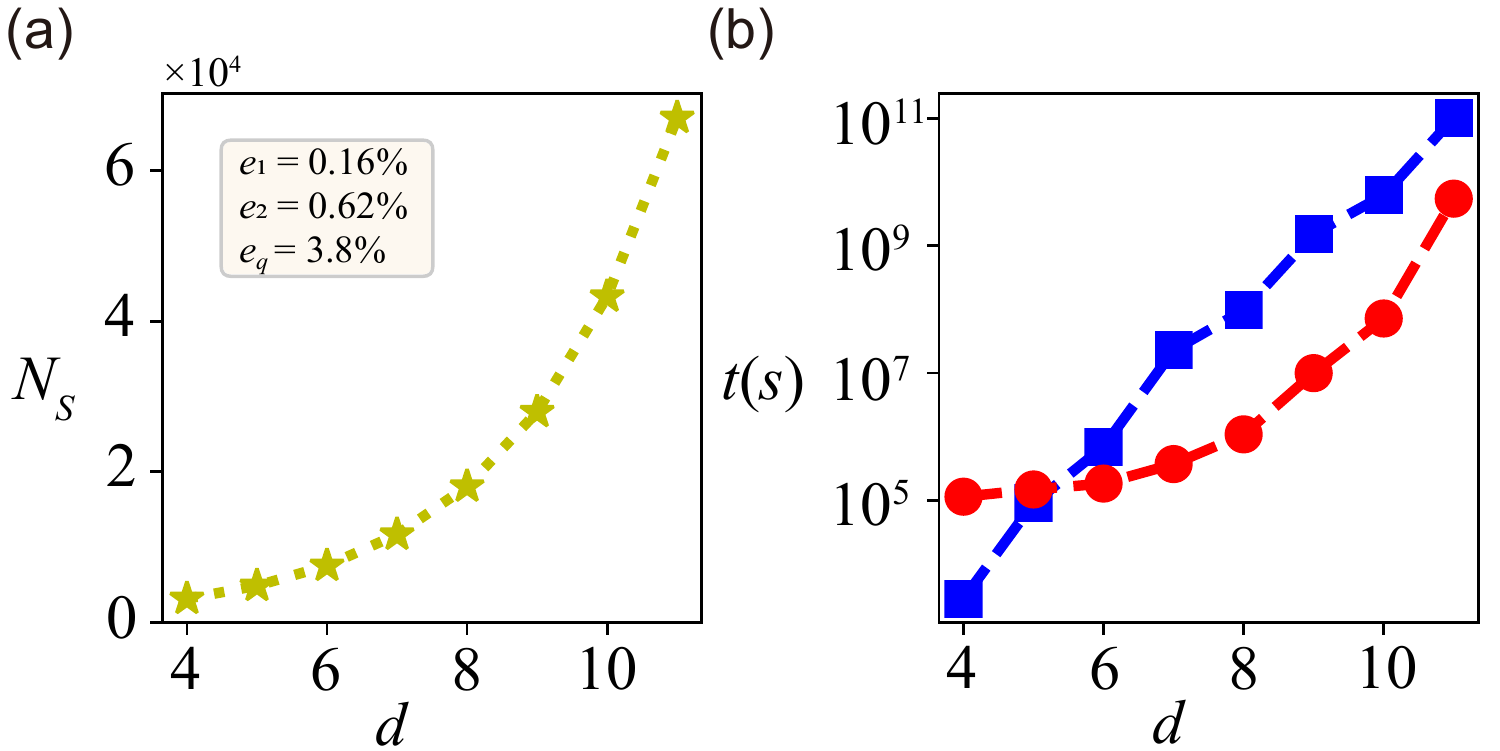}
\caption{(a) Smallest number of required bitstrings $N_s$ as a function of circuit depth $d$ for gate fidelities shown in the text box. (b) Estimated run time $t$ as a function of depth $d$. The blue dashed line with square and the red dashed line with circle are results for the Sch$\ddot{\text{o}}$dinger-Feynman algorithm and our TNS based algorithm respectively.}   \label{fig:fig3}
\end{figure}

\textit{Verification of RQCs.} We demonstrate the efficiency of our algorithm by applying it to simulate RQCs running on a 53-qubit Sycamore processor, and then comparing its performance to the Sch$\ddot{\text{o}}$dinger-Feynman algorithm~\cite{MarkovBoixo2018}. Our TNS based algorithm is a \textit{single-amplitude} algorithm since the complexity of computing $M$ amplitudes is equal to $M$ times the complexity of computing a single amplitude. In contrast, the Sch$\ddot{\text{o}}$dinger-Feynman algorithm is a \textit{full-amplitude} algorithm since computing a single amplitude is almost as hard as computing a bunch of $M$ amplitudes. Therefore for a fair comparison one needs to specify the smallest number of bitstrings $N_s$ required, for example, for the verification task. $N_s$ will in general increase as the fidelity $\fidelity$ of the quantum circuit decreases, which can be computed as
\begin{align}\label{eq:fidelity}
\fidelity = \prod\limits_{{\rm{g}} = {G}} {(1 - {e_{\rm{g}}})} \prod\limits_{{\rm{q}} = Q} {(1 - {e_{\rm{q}}})}.
\end{align}
Here $G$ denotes the gate set, $e_g$ denotes the gate error rate, $Q$ denotes the qubit set and $e_q$ denotes readout error rate. The Sycamore processor used in Ref.~\cite{AruteMartinisQuantumSupremacy2019} has a single qubit error rate of $e_1=0.16\%$, two-qubit gate error rate of $e_2=0.62\%$, and readout error rate of $e_q=3.8\%$. To ensure that $\fidelity$ is larger than $0$ with $3\sigma$, where $\sigma=1/\sqrt{N_s}$ denotes the statistical error, $N_s$ needs to satisfy $N_s \geq (3/F)^2$. We plot $N_s$ as a function of the circuit depth $d$ in Fig.~\ref{fig:fig3}(a).

In Fig.~\ref{fig:fig3}(b), we plot the estimated total run time $t$ for both algorithms as a function of $d$, where the blue dashed with square stands for the Sch$\ddot{\text{o}}$dinger-Feynman algorithm while the red dashed line with circle stands for our tensor network based algorithm. For the Sch$\ddot{\text{o}}$dinger-Feynman algorithm, we measure the time $t_0$ for a single path and then the total run time $t$ can be computed as $t = t_0N_p$ where $N_p$ is the total number of paths. For the TNS based algorithm, we compute the time $t_s$ for a single amplitude and then the total run time $t$ is simply $t=N_st_s$. Both simulations are done using a single thread of the Intel-Xeon-Gold-6254 CPU ($3.1$ GHz). We note that our native implementation of the Sch$\ddot{\text{o}}$dinger-Feynman algorithm has a performance similar to the record in Ref.~\cite{AruteMartinisQuantumSupremacy2019}, which is however still $1$ to $2$ orders of magnitudes slower than our TNS based algorithm for depths $6\leq d\leq 11$. Concretely, our TNS based algorithm is about $156$ times faster at $d=9$, and $19$ times faster at $d=11$. This would lead to significant cut down of the verification time. More details about the simulations down with both algorithms are in the supplementary~\cite{supp}.

We show the performance of our TNS based algorithm as the number of qubits $N$ increases in TABLE.~\ref{tab:benchmarktns}. We can see that for relatively shallow circuits (especially when $d\leq 8$), our algorithm is much more preferable than the Sch$\ddot{\text{o}}$dinger-Feynman algorithm. For example for $d=6$, the run time with our algorithm increases very little as $N$ increases from $54$ to $72$, while for the Sch$\ddot{\text{o}}$dinger-Feynman algorithm it will be significantly more difficult (at least by a factor of $2^9$) since one has to store and manipulate two $36$-qubit sub circuits \footnote{The increase in the number of simulated paths could also greatly increase the computational complexity of the Sch$\ddot{\text{o}}$dinger-Feynman algorithm (see Ref.~\cite{zlokapa2020boundaries} for details), which is not discussed here.}. Moreover, our algorithm can simulate the $104$-qubit RQC of Sycamore-like structure with up to $8$ depth, which is not possible for the Sch$\ddot{\text{o}}$dinger-Feynman algorithm simply due to the memory limitation (One has to store at least one $52$-qubit quantum state exactly)\footnote{Here we consider only the Sch$\ddot{\text{o}}$dinger-Feynman algorithm with 2 patches. Using more patches can reduce the memory consumption, but the time consumption may increase dramatically (see Ref.~\cite{zlokapa2020boundaries} for details).}.


\begin{table}
\caption{{\bf Run time to compute a single amplitude using TNS based algorithm. } RQCs on Sycamore-like structures of sizes $54,60,66,72,104$ and depths from $6$ to $10$ are simulated using a single thread. The run time is shown in seconds. NA means that data is not available. }
\label{tab:benchmarktns}
\centering
\begin{tabular}{| c | c | c | c | c | c |}
\Xhline{1pt}
\diagbox{Processor}{depth}     & 6 & 7 & 8 & 9 & 10 \\
\hline
Sycamore-54    & 24   & 19 & 143   & 1370 & 7260  \\
Sycamore-60    & 24   & 30 & 337   & 4145 & 53007  \\
Sycamore-66    & 29   & 82 & 1525   & 28968 & 267802  \\
Sycamore-72    & 41   & 465 & 21669   & 278679 & NA  \\
Sycamore-104    & 107   & 15177 & 458576   & NA & NA  \\
\Xhline{1pt}
\end{tabular}
\end{table}

In summary, we have presented a tensor network states based algorithm designed to simulate random quantum circuits with arbitrary geometry. We use singular value decomposition together with a two-sided circuit evolution algorithm to compress the size of the resulting tensor network from computing a single amplitude, which is further split into many smaller sub tensor networks using the \textit{cut} technique. We then propose a heuristic algorithm to find the optimal tensor contraction path. We demonstrate with numerical simulations that our algorithm is of $1$ to $2$ orders of magnitudes faster than the Sch$\ddot{\text{o}}$dinger-Feynman algorithm when simulating random quantum circuits on the $53$-qubit Sycamore processor for depths $6\leq d\leq 11$, and show that for relatively shallow RQCs our algorithm has a much more preferable scaling than the Sch$\ddot{\text{o}}$dinger-Feynman algorithm as the number of qubits increases. Therefore we expect that our algorithm could be the method of choice for the fast verification of NISQ hardwares.



\begin{acknowledgments}
C. G. acknowledges support from National Natural Science Foundation of China under Grants No. 11805279. H.-L. H. is supported by the Open Research Fund from State Key Laboratory of High Performance Computing of China (Grant No. 201901-01), NSFC (Grants No. 11905294), and China Postdoctoral Science Foundation.
\end{acknowledgments}


%

\end{document}


\title{Supplemental Material:\\Verifying Random Quantum Circuits with Arbitrary Geometry Using Tensor Network States Algorithm}

\date{\today}

\maketitle

\section{Heuristic algorithm to search for the optimal tensor contraction path}
In this section we present the details of our algorithm used to search for the optimal tensor contraction path. We denote an unordered collection of qubit indexes $i_1, i_2, \dots, i_m$ as $\{i_1, i_2, \dots, i_m\}$. A contraction path is specified by an ordered list of integers representing the corresponding qubit indexes, denoted as
\begin{align}\label{eq:path}
\cpath_{i_1i_2\dots i_m} = i_1 \rightarrow i_2 \rightarrow \dots \rightarrow i_m,
\end{align}
where $1\leq m\leq N$. When $m < N$, $\cpath_{i_1\dots i_m}$ denotes a partial tensor contraction path. $\cpath_{i_1i_2\dots i_m}$ is order-sensitive so that $\cpath_{i_1i_2}$ and $\cpath_{i_2i_1}$ represent different contraction paths. For each path $\cpath_{i_1 i_2\dots i_m}$, there is a corresponding group of tensors denoted as $\{C^{i_1}, C^{i_2}, \dots, C^{i_m} \}$. Contracting all the tensors in this group, we will get a resulting tensor denoted as
\begin{align}
C^{i_1 i_2\dots i_m} = \contract(C^{i_1}, C^{i_2}, \dots, C^{i_m}),
\end{align}
where the function $\contract(A_1, A_2, \dots, A_n)$ means to contract all the pairs of connected auxiliary indexes of the input tensors $A_1, A_2, \dots, A_n$. The rank of $C^{i_1 \dots i_m}$ is equal to the number of unconnected auxiliary indexes. When $m=N$, all the auxiliary indexes have been contracted with each other and $C^{i_1 \dots i_N}$ is a scalar. We can see that $C^{i_1 \dots i_N}$ is unique for a fixed group of qubit indexes, namely $C^{i_1 \dots i_m} = C^{i_1' \dots i_m'}$ as long as $\{i_1, \dots, i_m\} = \{i_1', \dots, i_m'\}$.

For each $\cpath_{i_1\dots i_m}$, we associate a score $\score_{i_1\dots i_m}$ which is a scalar representing the computational complexity of this path. We set $\score_{i_1}=0$ when $m=1$. When $\cpath_m$ absorbs a new qubit index $i_{m+1}$, $\score$ grows correspondingly
\begin{align}
\score_{i_1\dots i_m i_{m+1}} = \score_{i_1\dots i_m } + \Cost(C^{i_1 i_2\dots i_m}, C^{i_{m+1}}),
\end{align}
where the function $\Cost(A, B)$ means the computational complexity of contracting the two input tensors $A$ and $B$. Therefore the goal is to find the optimal tensor contraction path reduces to finding the path $\cpath_{i_1\dots i_N}$ with the least score $\score_{i_1\dots i_N}$. In the following we denote the optimal tensor contraction path as $\optpath_{\{i_1,\dots, i_N\}}$, and the corresponding score as $\optscore_{\{i_1,\dots, i_N\}}$, which satisfy
\begin{align}
\optscore_{\{i_1,\dots, i_N\}} = \min_{i_1\dots i_N \in \perm(N)} \score_{i_1\dots i_N}.
\end{align}
Here we have used $\{i_1,\dots, i_N\}$ in the subscripts of $\optpath$ and $\optscore$ to explicitly indicate that they are independent of the order of the underlying qubit indexes, and the function $\perm(N)$ means the group of all the possible permutations of the list $\{1,2,\dots, N\}$.

Directly searching for $\optpath_{\{i_1\dots i_N\} }$ by traversing over whole configuration space is apparently impossible due to the exponential growth of the number of possible contraction paths. For example, for the $53$-qubit Sycamore lattice there are a total of $53!$ paths. However the geometrical configurations of real quantum hardwares are in general not arbitrary, but instead have (quasi)-regular two-dimensional structures, such as the square lattice~\cite{VillalongaMandra2019}, the Sycamore lattice~\cite{AruteMartinisQuantumSupremacy2019}. As a result, $\optpath$ is should grow \textit{continuously}, that is, the successive qubit indexes $i_{n}$ and $i_{n+1}$ should be neighbours, since otherwise $\Cost{C^{i_1i_2\dots i_n}, C^{i_{n+1}}}$ would be very large since no pairs of auxiliary indexes will be contracted in this step. Since each qubit only has a few neighbours (at most $4$ for the above mentioned lattices), one may exclude a huge number of contraction paths.
To be concrete, we make the following assumptions:
\begin{enumerate}
\item There are well-defined boundaries for the lattice configurations of real quantum hardwares which has a fewer number of neighbours, and paths starting from those boundary qubits have less scores; \label{cond:cond1}
\item The rank of the tensor $C^{i_1\dots i_m}$ appeared along the optimal path should be less than a maximal value $\edgemax$ for all $1\leq m\leq N$. \label{cond:cond2}
\item The partial contraction path $\cpath_{i_1\dots i_m}$ is itself either a connected graph, or a connected graph plus a single isolated qubit index. If the latter case happens, then $i_{m+1}$ must be such that $\cpath_{i_1\dots i_m i_{m+1}}$ is connected; \label{cond:cond3}
\end{enumerate}
Assumption.~\ref{cond:cond1} eliminates the paths starting from interior qubits and Assumption.~\ref{cond:cond2} prohibits the occurrence of very high-dimensional intermediate tensors. $\edgemax$ could be chosen as the tree width of the underlying graph, for example. Assumption.~\ref{cond:cond3} eliminates the cases that there are two or more disconnected paths with different starting points, which grow and finally join each other. There is no guarantee that latter approach is more expensive than the case of a single main path. However in cases of quasi-regular lattices, one could often find a single main contraction path with a similar numerical complexity to the case of several disconnected contraction paths, especially if parallelization is not considered, which is indeed the case here since it is much more efficient and elegant to parallelize this program on the level of computing several amplitudes or contracting those sub tensor networks encountered when computing a single amplitude. We allow a single isolated qubit simply because for the current quantum computing hardwares there may exist qubits with a single neighbour, for such cases and for certain paths adding such an isolated qubit may be more beneficial in terms of the size of the resulting tensor. These three assumptions, especially the last one, allow us to filter out most of the tensor contraction paths.


In the following we present an efficient searching algorithm based on the state compression dynamical programming technique and then give a justification for it. We denote all the necessary information attached with a partial contraction path $\cpath_{i_1\dots i_m}$ as a three-tuple $\pathinfo_{i_1\dots i_m} = \{\cpath_{i_1\dots i_m}, \score_{i_1\dots i_m}, c \}$, where $c$ is an additional integer indicating the connectivity of $\cpath$. If $\cpath$ is connected then $c = -1$, otherwise $c$ is a positive integer representing the index of the isolated qubit. We use the priority queue data structure to store all the possible $\pathinfo$s, denoted as $\allpathinfo = \{\pathinfo_{i_1 \dots i_m}, \pathinfo_{j_1 \dots j_n}, \dots\}$. $\allpathinfo$ supports two operations $\push$ and $\pop$, where the function $\push(\allpathinfo, \pathinfo_{i_1\dots i_m})$ inserts a new item $\pathinfo_{i_1\dots i_m}$ into $\allpathinfo$, while the function $\pop(\allpathinfo)$ means to take the item $\pathinfo$ with the least score $\score$ out of $\allpathinfo$. We use $\nodes^B$ to denote the collection of the indexes of the qubits on the boundaries.

The algorithm is shown in Algorithm.~\ref{alg:alg1}. We have also made use of two more auxiliary functions $\neighbour$ and $\conn$. The function $\neighbour(\cpath_{i_1\dots i_m})$ outputs all the possible new qubit indexes satisfying Assumptions.~(\ref{cond:cond2}, \ref{cond:cond3}). The function $\conn(\cpath_{i_1\dots i_m})$ outputs an integer $c$ representing the connectivity of the path $\cpath_{i_1\dots i_m}$, that is, $c=-1$ if there is at least one edge between $i_{m+1}$ and any of $i_1, \dots i_m$ and $c=i_{m+1}$ otherwise. Assumption.~\ref{cond:cond1} is used in the initialization stage, where we restrict ourself to start from the boundary qubits. Assumptions.~(\ref{cond:cond2}, \ref{cond:cond3}) are used when using the function $\neighbour$.

\begin{algorithm}[H]
\DontPrintSemicolon
\SetAlgoLined
\KwResult{Optimal tensor contraction path $\optpath_{\{i_1,\dots, i_N\} }$}
 Initialize $\set = \{\}$, $\allpathinfo = \{\pathinfo_{i_1}, \pathinfo_{j_1},\dots \}$ with $i_1, j_1,\dots \in \nodes^B$ and $\pathinfo_{i_1} = \{\cpath_{i_1}, 0, -1\}$, $\pathinfo_{j_1} = \{\cpath_{j_1}, 0, -1\}$, $\dots$\;

\While{true}{
	$\pathinfo_{i_1\dots i_m} = \pop(\allpathinfo) $\;
	\If{$\{i_1, \dots, i_m\} \in \set$}{\Continue}
	$\push(\set, \{i_1, \dots, i_m\})$\;
	\If{$m == N$}{\Return{$\cpath_{i_1\dots i_m}$}}
	\For{$i_{m+1} \in \neighbour(\cpath_{i_1\dots i_m})$}{
		$\score_{i_1\dots i_m i_{m+1}} \leftarrow \score_{i_1\dots i_m} + \Cost(\cpath_{i_1\dots i_{m}}, i_{m+1}) $\;
		$c \leftarrow \conn(\cpath_{i_1\dots i_m i_{m+1}})$\;
		$\pathinfo_{i_1\dots i_m i_{m+1}} = \{ \cpath_{i_1\dots i_m i_{m+1}}, \score_{i_1\dots i_m i_{m+1}}, c \}$\;
		$\push(\allpathinfo, \pathinfo_{i_1\dots i_m i_{m+1}})$\;
	}
}

 \caption{Algorithm to search for the optimal tensor contraction path.} \label{alg:alg1}
\end{algorithm}

Now we give a justification of Algorithm.~\ref{alg:alg1}. An important observation is that if $\cpath_{i_1\dots i_m}$ is the optimal contraction path among all the possible paths for a fixed set of qubit indexes $\{i_1, \dots, i_{m-1}, i_m\}$, then $\cpath_{i_1\dots i_{m-1}}$ must also be the optimal contraction path for the fixed set of qubit indexes $\{i_1, \dots, i_{m-1}\}$. This can be proven straightforwardly as follows: for any $\cpath_{i_1'\dots i_{m-1}'}$ with $\score_{i_1'\dots i_{m-1}'} > \score_{i_1\dots i_{m-1}}$, where $\{i_1'\dots i_{m-1}' \}$ is a permutation of $\{i_1\dots i_{m-1} \}$, we will have
\begin{align}
\score_{i_1'\dots i_{m-1}' i_m} &= \score_{i_1'\dots i_{m-1}'} + \Cost(C^{i_1'\dots i_{m-1}'}, C^{i_m}) \nonumber \\
&> \score_{i_1\dots i_{m-1}} + \Cost(C^{i_1'\dots i_{m-1}'}, C^{i_m}) \nonumber \\
&= \score_{i_1\dots i_{m-1}} + \Cost(C^{i_1\dots i_{m-1}}, C^{i_m}) \nonumber \\
&= \score_{i_1\dots i_{m-1} i_m} ,
\end{align}
where we have used the fact that $C^{i_1\dots i_{m-1}}$ and $C^{i_1'\dots i_{m-1}'}$ are the same tensor. Then we have
\begin{align}\label{eq:optscore}
\optscore_{\{i_1,\dots, i_m\} } = \min_{i_m' \in \{i_1,\dots, i_m\} }\left(\optscore_{\{i_1',\dots, i_{m-1}'\} } + \Cost(C^{i_1'\dots i_{m-1}'}, C^{i_m'}) \right).
\end{align}
Since $\optscore_{\{i_1', \dots, i_{m-1}' \} }$ and $C^{i_1'\dots i_{m-1}'}$ are both independent of the order of $\{i_1'\dots i_{m-1}'\}$, there are only $m$ possible candidates on the right hand side of Eq.(\ref{eq:optscore}). Eq.(\ref{eq:optscore}) means that to find $\optpath_{\{i_1,\dots, i_m \} }$, one only needs to find all the $m$ possible $\optpath_{\{i_1',\dots, i_{m-1}' \} }$s. Therefore without Assumptions.(\ref{cond:cond1}, \ref{cond:cond2}, \ref{cond:cond3}), namely in the initialization stage and the $\neighbour$ function we allow the most generic case, then Algorithm.~\ref{alg:alg1} is guaranteed to find the global optimal contraction path. In practice, we find that when these contrains are imposed, Algorithm.~\ref{alg:alg1} returns a reasonably well contraction path at least for current quantum hardware geometries.

\section{Details for the numerical simulations}

\begin{table}[!htb]
\centering
\caption{{\bf Comparison between our native implementation of the Sch$\ddot{\text{o}}$dinger algorithm with the data in Ref.~\cite{AruteMartinisQuantumSupremacy2019}.} In our simulations we have used four CPUs (Intel-Xeon-Gold-6254, 3.1GHz) with 18 cores each, while in Ref.~\cite{AruteMartinisQuantumSupremacy2019} they have used n1-ultramem-160, which has four CPUs (2.2 GHz) with $20$ cores each. NA means that data is not available.}
\label{tab:benchmarksa}
\begin{tabular}{| p{2cm} | p{2.5cm} | p{2.5cm} |}
\Xhline{1pt}
Num. of qubits     & run time in seconds (ours) & run time in seconds (Google) \\
\hline
30    & 28   & NA  \\
32                   & 93   & 111   \\
34                   & 362 & 473   \\
36                   & 1357 & 1954   \\
\Xhline{1pt}
\end{tabular}
\end{table}

We use the Sch$\ddot{\text{o}}$dinger-Feynman algorithm as the benchmarking baseline for our numerical simulations. In the Sch$\ddot{\text{o}}$dinger-Feynman algorithm, the full $53$-qubit circuit is split into two sub circuits of $26$ and $27$ qubits separately. Then those two sub circuits are evolved independently until a \textit{cross} gate $O$ acting on both groups is met, in which case the evolution is split into $\chi_o$ independent paths weighted by the $\chi_o$ singular values if $O$ as defined in the main text. For detailed descriptions of this algorithm one can refer to Refs.~\cite{MarkovBoixo2018,AruteMartinisQuantumSupremacy2019}. Therefore given a specific way of splitting the full circuit, the total number of paths is determined by the number of cross gates (denoted as $n_c$) as well as the rank $\chi_o$ of them, which is $\chi_o^{n_c}$.

In practice, one can make a \textit{checkpoint} at a particular cross gate by saving a copy of the states corresponding to those two sub circuits, and then for rest evolution of each path could start from those copies instead of evolving from the initial states. The total number of gate operations could be significant reduced by this technique, the price to pay is an additional copy of each quantum state of the sub circuits, which is manageable at a scale of $26$ to $27$ qubits. With this technique, assuming there are $n_p$ cross gates before this checkpoint, then the total number of paths is counted as $N_p = \chi_o^{n_p}$ instead, as done in~\cite{AruteMartinisQuantumSupremacy2019}. We implement the Sch$\ddot{\text{o}}$dinger-Feynman algorithm natively, in which the checkpoint $n_p$ is chosen such that the total number of gate operations is minimized. In particular, we have chosen $n_p=3,4,7,8,11,11,14$ for $d=5,\dots, 11$ respectively. Moreover, since the performance of the Sch$\ddot{\text{o}}$dinger-Feynman algorithm is ultimately determined by the performance of the Sch$\ddot{\text{o}}$dinger algorithm on each sub circuit, here we show the performance of our Sch$\ddot{\text{o}}$dinger algorithm based simulator in in TABLE~\ref{tab:benchmarksa}, from which we are confident that our implementation of the Sch$\ddot{\text{o}}$dinger-Feynman algorithm has at least the same level of performance compared to Ref.~\cite{AruteMartinisQuantumSupremacy2019}.

\begin{figure}[tbp]
\includegraphics[width=\columnwidth]{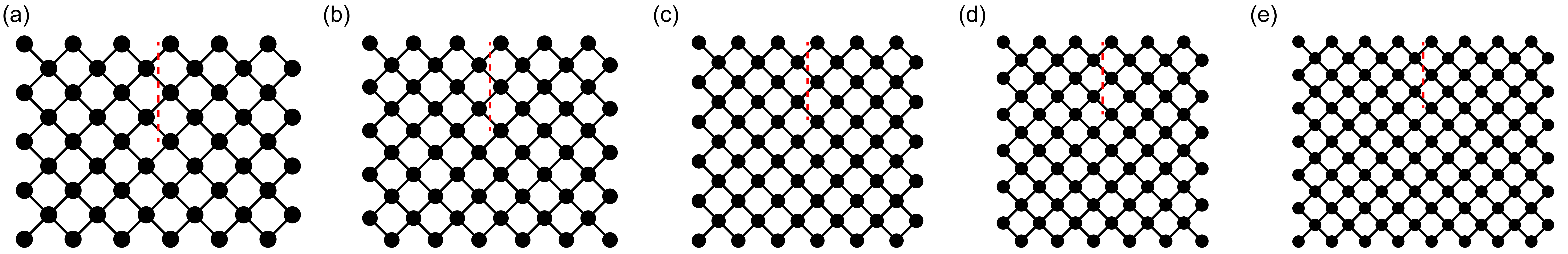}
\caption{Quantum processors of Sycamore-like structure with different sizes used in the main text for (a) 54-qubits, (b) 60-qubits, (c) 66-qubits, (d) 72-qubits, (e) 104-qubits. Each circle represents a qubit and the edges between the qubits represent the positions of the two-qubit gate operations. The vertical dashed line represents the positions of the cuts which have been used in the simulation results from our tensor network states based algorithm. }   \label{fig:figs1}
\end{figure}

For the tensor network states based algorithm, we have used the \textit{cut} technique to further split the resulting tensor network from computing a single amplitude into many smaller sub tensor networks. After that, we compute the best tensor contraction path $\optpath$ for one of those sub tensor networks using Algorithm.~\ref{alg:alg1}, and then contract all those sub tensor networks along this path. The positions as well as the number of cuts are chosen empirically, which are shown in Fig.~\ref{fig:figs1}. As a rule of thumb, one should cut the edges where the lattice is the thinnest.

